\documentclass{article}
\usepackage{ijcai24}

\usepackage{times}
\usepackage{soul}
\usepackage{array}
\usepackage{url}
\usepackage[hidelinks]{hyperref}
\usepackage[utf8]{inputenc}
\usepackage[small]{caption}
\usepackage{graphicx}
\usepackage{subcaption}
\usepackage{amsmath}
\usepackage{amsthm}
\usepackage{amssymb}
\usepackage{booktabs}
\usepackage{algorithm}
\usepackage[switch]{lineno}
\usepackage{algpseudocode}

\title{Evaluating Co-Creativity using Total Information Flow}

\author{
Vignesh Gokul$^{*}$
\and
Chris Francis$^{*}$\And
Shlomo Dubnov
\affiliations
UC San Diego\\
$^*$ denotes equal contribution\\
\emails
\{vgokul, cfrancis, sdubnov\}@ucsd.edu,
}

\begin{document}

\maketitle

\begin{abstract}
Co-creativity in music refers to two or more musicians or musical agents interacting with one another by composing or improvising music. However, this is a very subjective process and each musician has their own preference as to which improvisation is better for some context. In this paper, we aim to create a measure based on total information flow to quantitatively evaluate the co-creativity process in music. In other words, our measure is an indication of how ``good" a creative musical process is. Our main hypothesis is that a good musical creation would maximize information flow between the participants captured by music voices recorded in separate tracks. We propose a method to compute the information flow using pre-trained generative models as entropy estimators. We demonstrate how our method matches with human perception using a qualitative study.
\end{abstract}

\section{Introduction}
Human creativity is a complex process that stems from a wide range of experiences, complex thoughts, and emotions. The goal of objectively evaluating human creativity has been a long-standing research problem across the fields of psychology and cognitive science. Most of the existing works \cite{brunner2018symbolic,cifka2020groove2groove,dubnov2022creative,dubnov2010information} attempt to evaluate creativity in specific works of art such as music, visual art, etc. This is a complex problem as creativity is an abstract concept that even we humans do not understand completely. Humans are also known to co-create, i.e. participate in the creation process together by assisting each other. Some examples of co-creativity spans the fields of music \cite{dubnov2022creative,dong2018musegan,jiang2020counterpoint} (musical improvisation), arts (paintings) \cite{rombach2022high} etc. With the rise of artificial generative models, humans have started co-creating along with agents as well in different art forms. This has led to an interest in evaluating co-creativity between agents and humans, which is not a simple task.
\begin{figure}[t]
    \centering
    \includegraphics[width = 0.5\textwidth]{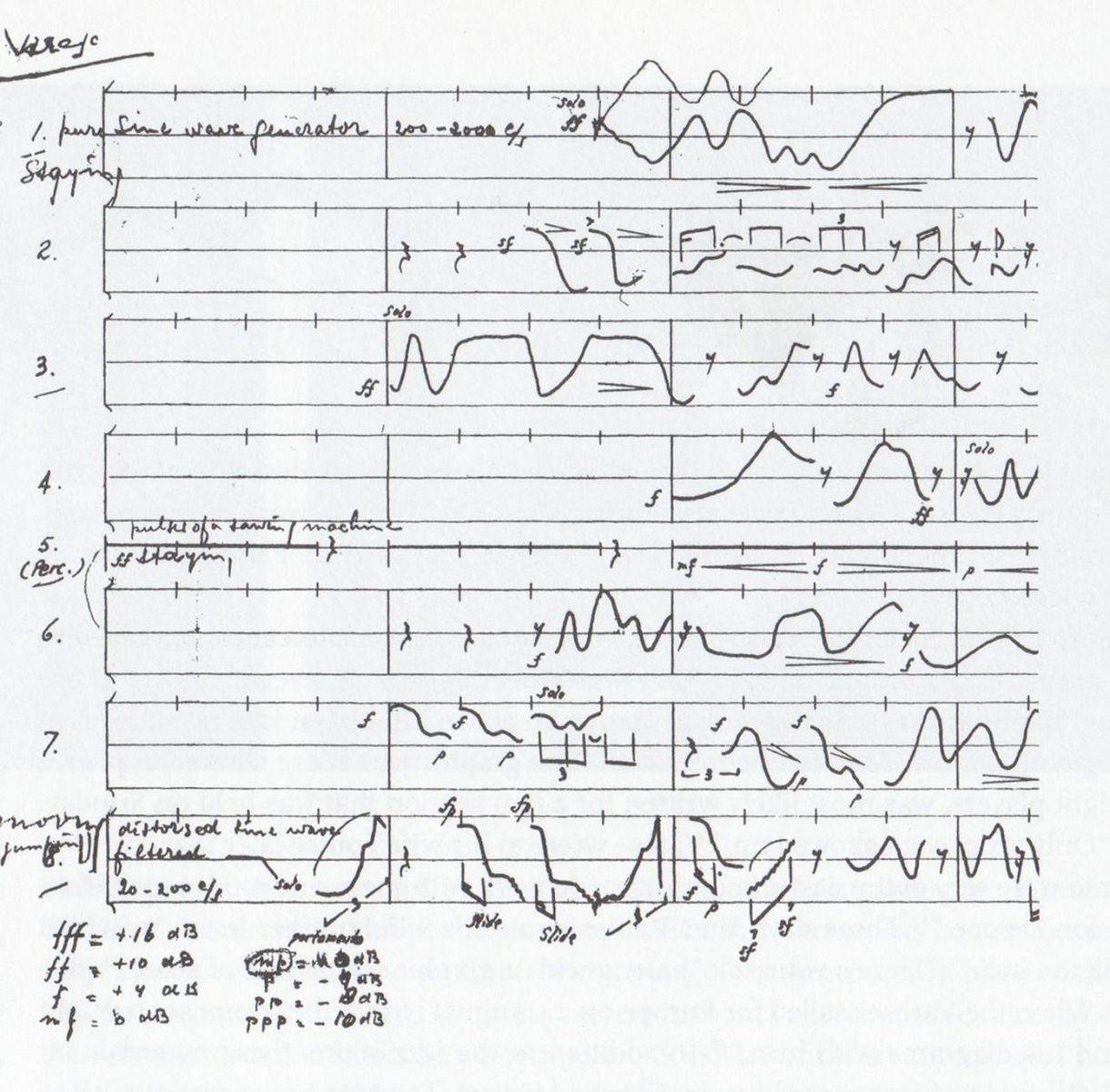}
    \caption{Excerpt from an electronic piece ``Poeme Electronique" by Edgar Varese}
    \label{fig:poeme}
\end{figure}
Generative models have shown remarkable potential in matching humans in creative arts. This includes tasks like image generation (painting, drawing, etc) \cite{rombach2022high}, video generation (movie making) \cite{brooks2022generating} and audio generation \cite{mmt,anticipatory} (music, singing, and speech). While most of the frameworks focus on autoregressive sequence generation i.e., continual generation from an input sequence, recent works \cite{gokul2019semantic,dubnov2022creative,dubnov2023switching} have tackled the problem of interactive creation: an agent responds to the data rather than extending the data. In this work, we specifically focus on the problem of musical improvisation: a human and an agent/another human plays music together, responding to each other. 

Although there has been impressive research on building generative models for audio\cite{anticipatory,copet2023simple}, there is a severe lack of evaluative measures to understand the creative process behind the generated data. Understanding how creativity works in humans is essential to mimic the same behavior in artificial agents. For example, if we can come up with a score that quantifies this creative process, then attempts can be made to use the score to guide the model to mimic human creativity. But this is not a trivial task. The quality of musical improvisation is difficult to assess. It is a subjective evaluation, as many consider musical improvisation to be an expression of personal emotions. There is no ``right" or ``wrong" music. This leads to an interesting question: can we quantify this creativity preference using a score? A possible hint towards an answer can be drawn from analyzing an abstract score as shown in Figure \ref{fig:poeme} \footnote{ Excerpt from an electronic piece ``Poeme Electronique" by Edgar Varese}. In this piece, the music comprises multiple sonic materials notated on different lines, with pitch represented by horizontal curves, rhythmic elements as vertical stems, and additional angled lines specifying gradual changes in loudness. According to our approach, the total music information in this piece is ``co-created" via interaction across multiple voices according to their individual and codependent dynamics.  

Most of the existing research attempts to quantify this creative process involve some form of subjective studies\cite{agarwala2017music,hadjeres2017deepbach,huang2019counterpoint}. This involves asking participants to evaluate the score on different aspects and using statistical measures on the responses to come up with a final measure. Recent works have attempted to use information theory measures such as transfer entropy and mutual information to evaluate musical improvisation/accompaniment. While these measures are a good approximation, they are extremely hard to calculate on high-dimensional data such as music and audio.
\begin{figure}[t]
    \centering
    \includegraphics[width = 0.5\textwidth]{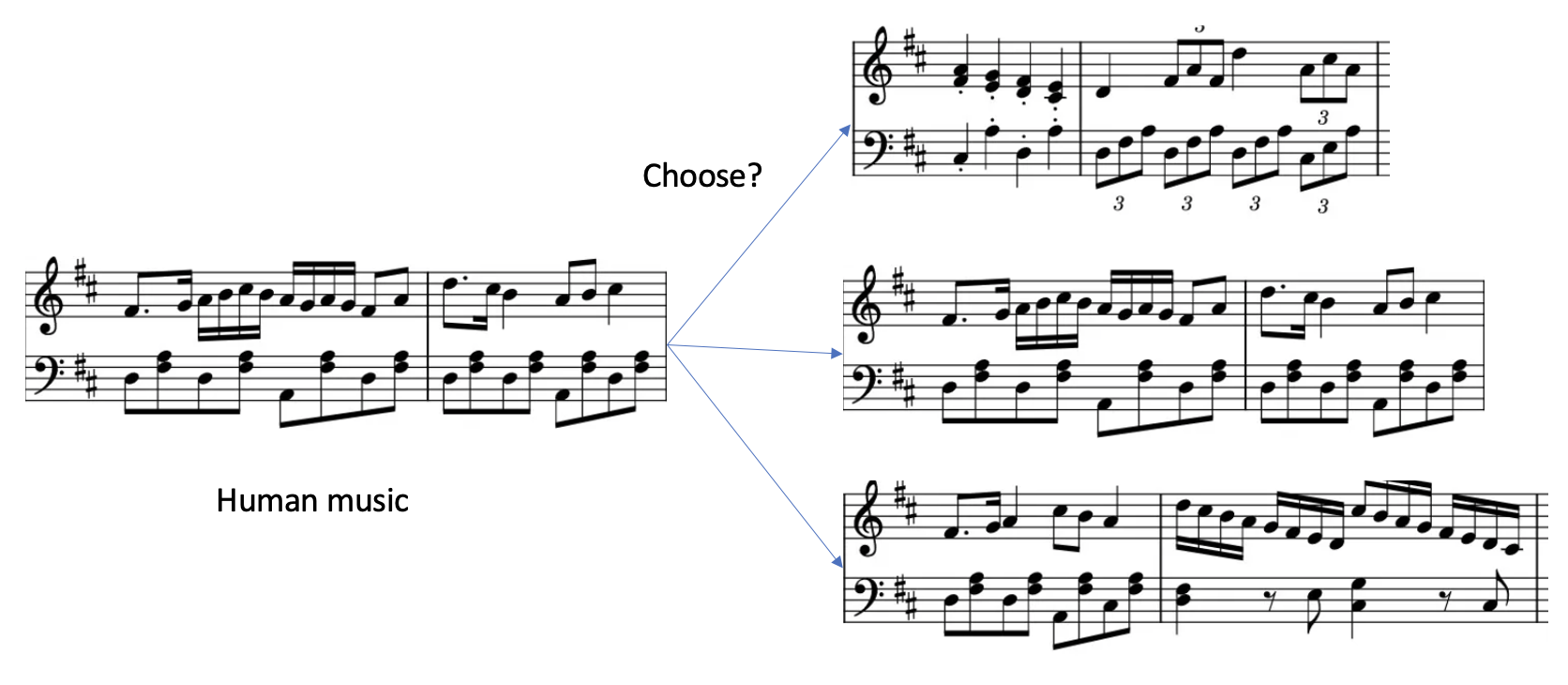}
    \caption{For a given input musical signal from a human musician, what represents the best corresponding improvisation? The response must be creative and semantically make sense.}
    \label{fig:intro}
\end{figure}
We aim to create an evaluation measure based on the amount of musical information passed between the two musical signals. Our hypothesis is that, for the ``most creative" signal, the amount of information flowing from the human to the agent and vice versa is maximized. This is based on the fact that musical improvisation is essentially a form of non-verbal communication. Each musician tries to convey/express some information while playing and similarly responds to the other agent/human's music by expressing data. In other words, we hypothesize that \emph{the amount of information flow between two musicians is an indication of creativity.} For example, as shown in Figure \ref{fig:intro}, given an input musical piece, which of the three possible outputs represents the best musical improvisation output? Based on our hypothesis, the musical score maximizing the information flow would be the best choice.
In this paper, we present a method to calculate the amount of information flow between two musical processes. We leverage pre-trained generative models as entropy approximators to make computing the score tractable for long musical sequences. Our analysis and subjective studies show that our score matches with human perception of creativity.

Our contributions are summarized as follows: 1) We propose an evaluation score to quantify the co-creativity process using directed information. 2) We design a framework to compute the score using pre-trained generative models. 3) Finally, we demonstrate the effectiveness of our score with an extensive analysis on multiple datasets and subjective studies. 
\section{Related Work}
\subsubsection{Creativity} Research on precisely defining human creativity has spiked interest for a long period of time. Many have attempted to define what exactly constitutes creativity in the past \cite{plucker2004isn,simonton2018defining,taylor19884,treffinger1993stimulating}. \cite{runco2012standard}
define creativity as something that is novel and useful. Some definitions include surprisal \cite{simonton2018defining} and discovery of new possibilities \cite{martin2017defining}, while most of the works still refer to novelty and usefulness. 

\subsubsection{Co-Creativity in music}
Using computer systems in music production is not new. \cite{biles1994genjam,marques2000music}. With the rapid development of AI models and tools, the usage of artificial agents/programs in music composition, editing, and performance has been on the rise. While most of the recent works lean towards music generation \cite{mmt}, music inpainting \cite{moliner2023diffusion}, there have been works to enable co-creative generation tasks similar to music improvisation \cite{jiang2020counterpoint,jiang2020rl,wang2016machine}. \cite{jiang2020rl} propose a framework using Reinforcement Learning (RL) to enable real-time human-agent interaction. \cite{jiang2020counterpoint} introduce a mechanism to compute intrinsic rewards to guide counterpoint music generation. Although research has progressed a lot in building generative models, evaluating its creativity still remains a core problem in machine learning.
\subsubsection{Evaluating Creativity in Music}
The majority of the evaluation mechanisms in music stem from some form of subjective assessment \cite{agarwala2017music,hadjeres2017deepbach,huang2019counterpoint}. Turing tests \cite{turing2009computing} used to be a popular form of subjective assessment. However, it was not clear if the assessment was based on the quality of the generated music or whether the generated music was human-made. There has been a lot of research on the exact design of such assessments and user studies. \cite{dong2018musegan} analyzed visual aspects of the generated music, such as piano roll and spectrogram representations, to make conclusions on the melody flow and rhythm patterns. However, all these methods provide qualitative results, which are not great for consistent evaluation.  \cite{huang2019counterpoint} propose a quantitative measure based on the frame-wise negative log-likelihood between the generated music and ground truth. There have also been several music domain metrics proposed, such as pitch and rhythm-related \cite{yang2020evaluation}, harmony-related \cite{yeh2021automatic}, and style-related \cite{brunner2018symbolic,cifka2020groove2groove} metrics. Recently, measures that include utilizing deep learning models have become popular. Frechet Audio Distance (FAD) \cite{kilgour2018fr} provides a score based on statistical measures of the computed deep features of both the original and real audio.

However, most of these measures focus on evaluating the quality of the music and not the creativity involved in the music generation process. For example, to evaluate a music improvisation system, we can use FAD to evaluate the quality of the generated music (amount of noise, musical structures, etc), but it would not be possible to evaluate if the system is performing the ``right" type of improvisation for the input or if it is ``creative".

\subsubsection{Music Information Dynamics}
There have been several works to understand music structure by exploring the information content. Music Information Dynamics is a technique inspired by musical anticipation, i.e. quantifying the amount of information flow from past to future \cite{pearce2012auditory}.   \cite{abdallah2009information} use time-varying measures such as predictive information rate to characterize human perception of music.  \cite{dubnov2010information} calculate information rate at varying time scales using statistics of sound spectral features. \cite{dubnovdeep} propose Deep Music Information Dynamics. The authors estimated Mutual Information using Mutual Information Neural Estimators \cite{belghazi2018mine} (MINE) between two MIDI voices. They considered average MI between bars of music without consideration of temporal or causality relationships. In this work, we extend and reformulate the problem as information flow computation. 

\subsubsection{SymTE}
One of the closest works to our paper is \cite{dubnov2023switching}. The authors propose SymTE, a method to estimate information flow between two musical processes. The authors propose to switch to a new generative model based on which model outputs maximize the total information flow, which they call Symmetric Transfer Entropy (SymTE). Given two processes, $X$ and $Y$, the authors approximate the total information flow as:
\begin{align*}
    SymTE &\approx I((X,Y);\overline{(X,Y)}) - I(X,\bar{X})\\
    &- I(Y,\bar{Y}) + I(X,Y) 
\end{align*}
where $I$ is the predictive information. 
In order to compute the predictive information, they use the generative models in \cite{wang2016machine}. While SymTE shows promising directions in estimating total information flow and the potential of using such frameworks in generative modeling, it is not tractable to compute for long sequences like music. The framework we propose can easily be computed for long musical signals.

\section {Method}
The main objective of our work is to calculate a score to evaluate how semantically meaningful two musical signals are when played together. We consider two-track musical signals denoted by $(X,Y) = [x_i, y_i]$. Usually, in co-creative tasks like music improvisation, one signal (say $X$) comes from a human musician, while the other comes from an agent/another musician ($Y$). Our goal is to estimate the amount of information flow between both $X$ and $Y$. Given signals $(X,Y)$, we have:
\begin{align}
T_{X \to Y} &= I(Y;\bar{X}|\bar{Y}) \\
T_{Y \to X} &= I(X;\bar{Y}|\bar{X}) 
\end{align}
where $T_{X \to Y}$ represents the transfer entropy from $X \to Y$. Transfer entropy between two sequences is the amount of information passing from the past of one sequence to another when the dependencies of the past of the other sequence (the sequence's own dynamics) have already been taken into account. $\bar{X}$ denotes the past of the signal $X$.
We define the total information flow between $X$ and $Y$ as:
\begin{align*}
    \text{Flow} &= T_{X \to Y} + T_{Y \to X}
\end{align*}
 
Expanding on $T_{X \to Y}$, we have:
\begin{align*}
    T_{X \to Y} &= I(Y;\bar{X}|\bar{Y}) \\
&= H(Y|\bar{Y}) - H(Y|\bar{X},\bar{Y}) \\
&= H(Y|\bar{Y}) - H(Y|\bar{X},\bar{Y}) - H(Y) + H(Y) \\
&= I(Y;\bar{X},\bar{Y}) - I (Y;\bar{Y})
\end{align*}
Similarly, we have,
\begin{align*}
  T_{Y \to X} &= I(X;\bar{Y},\bar{X}) - I(X;\bar{X}) 
\end{align*}
From SymTE \cite{dubnov2023switching}, we know that
\begin{align*}
    T_{X \to Y} + T_{Y \to X} &\approx I((X,Y);\overline{(X,Y)}) - I(X,\bar{X})\\
    &- I(Y,\bar{Y}) + I(X,Y) 
\end{align*}

Expanding, we have \\
\begin{align*}
    T_{X \to Y} + T_{Y \to X} &= H(XY) - H(XY|\overline{XY}) + I(XY) \\
&- \left( H(X) - H(X|\bar{X}) \right) - \left( H(Y) - H(Y|\bar{Y}) \right) \\
\end{align*}

Since,
\begin{align*}
    I(X,Y) &= H(X) + H(Y) - H(XY),
\end{align*}
 we have, \\
\begin{align}
    \label{overall_info_flow}
    T_{X \to Y} + T_{Y \to X} &= H(X|\bar{X}) + H(Y|\bar{Y}) \\ \nonumber
    &- H(XY|\overline{XY})
\end{align}

In other words, the overall information flow between $X$ and $Y$ can be approximated by removing the surprisal of the combined signal $XY$ from the surprisal present in both $X$ and $Y$. For example, if the combined sequence $XY$ does not make sense musically, then it will have a higher entropy or suprisal factor, which means $H(XY|\overline{XY})$ will be higher, leading to lower total information flow. Similarly, if the signal $XY$ is musically pleasing, then ideally, it should be easier to predict the next note, i.e. low surprisal or entropy. This will cause the overall information flow in equation \ref{overall_info_flow} to be higher.

Note that the effectiveness of the score depends on the methods used to estimate each of the entropy terms. If we have poor entropy estimators then the values of the information flow would not align with our intuitive reasoning. Another important requirement is that our estimators should work for long musical sequences, which are high dimensional data. To summarize, we need strong, robust entropy estimators for high dimensional data to compute the terms in equation \ref{overall_info_flow}.

\begin{algorithm}
\caption{Calculate Information Flow}
\begin{algorithmic}[1]
\State \textbf{Input:} A MIDI file with two tracks (XY)
\State \textbf{Output:} Total Information Flow
\State $X, Y \gets \Call{SplitTracks}{XY}$ \Comment{Separate two tracks}
\For{each $T \in \{X, Y, XY\}$}
    \For{$i \gets 0$ \textbf{to} $seq\_len$}
        \State $start \gets T[i:\text{context}]$ 
        \State $probs \gets \Call{MTMT.Transformer}{start}$ 
        \State $cond\_entropy[T][i] \gets -\sum probs \times \log(probs)$ 
    \EndFor
    \State $cond\_entropy[T] \gets \Call{avg}{cond\_entropy[T]}$
\EndFor
\State $InfoFlow \gets cond\_entropy[X] + cond\_entropy[Y] - cond\_entropy[XY]$
\State \Return $InfoFlow$
\end{algorithmic}
\label{alg_1}
\end{algorithm}

\subsection{Estimating Entropy}
Computing the entropy terms in equation \ref{overall_info_flow} is a challenging task, especially when the data is high-dimensional such as music. We can leverage pre-trained generative models to approximate each entropy term. The conditional entropy refers to the amount of surprisal in a random variable $Y$, given that another random variable $X$ is known. We have:
\begin{align}
    H(Y|X) = - \sum_{x \in \mathcal{X}, y \in \mathcal{Y}} p(x, y) \log \frac{p(x,y)}{p(x)}
\end{align}
where $\mathcal{X}$ and $\mathcal{Y}$ denotes the support sets of $X$ and $Y$.
The conditional entropy can also be rewritten as:
\begin{align}
    H(Y|X) = \mathbb{E} \left [- \log \frac{p(x,y)}{p(x)} \right]
    &= \mathbb{E} \left [- \log p(y|x) \right]
\end{align}
Intuitively, the conditional entropy measures how unpredictable (amount of surprisal) $Y$ is on average, given that we have observed $X$. Now the conditional entropy given the past sequence becomes:
\begin{align}
    \label{cond_entropy}
    H(X|\bar{X}) = \mathbb{E} \left [- \log p(x|\bar{x}) \right]
\end{align}
Since we are dealing with temporal sequences (music audio), we take the expectation over time. To compute $p(x|\bar{x})$, we use pre-trained music generative models that give us the probability distribution given some context signal (past). Similarly, we also compute $H(Y|\bar{Y})$ and $H(XY|\overline{XY})$. 

\subsection{MIDI Transformer}
In this work, we use the Multitrack Music Transformer (MTMT) model \cite{mmt} pre-trained on the Symbolic Orchestral Database (SOD) dataset \cite{sod} to estimate entropy terms in equation \ref{cond_entropy}. The model is based on a decoder-only transformer architecture \cite{decoder1,decoder2} and has multi-dimensional input and output spaces similar to \cite{multi}. We use a learnable absolute positional encoding in the model \cite{transformer}. The loss function used is the sum of the cross-entropy losses of the six different fields in the data representation used by the model. 
\begin{figure*}[t] 
    \centering 
    \begin{subfigure}[b]{0.5\textwidth} 
        \centering
        \includegraphics[width=0.9\textwidth]{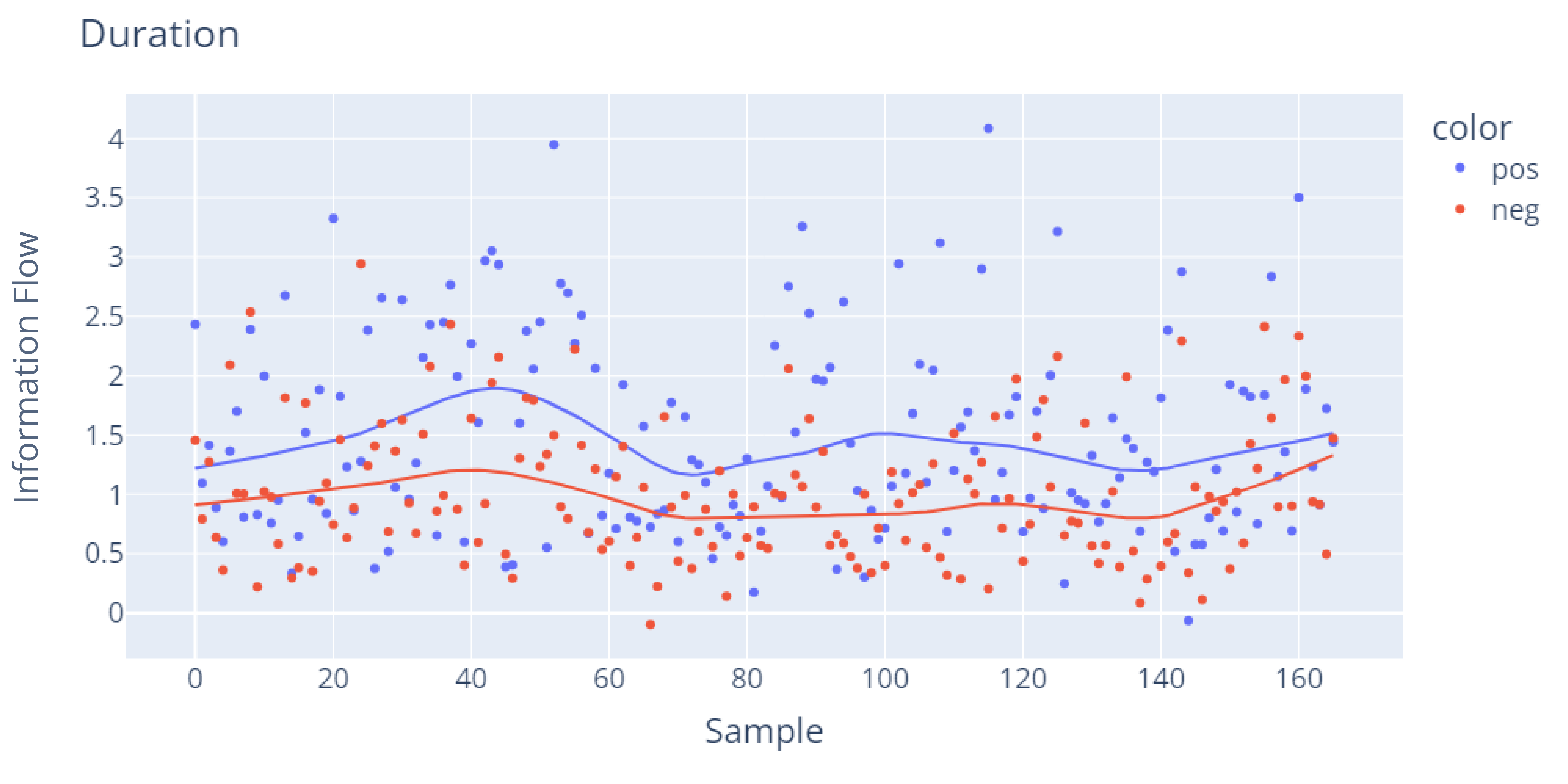} 
        \caption{}
        \label{fig:first}
    \end{subfigure}%
    \begin{subfigure}[b]{0.5\textwidth}
        \centering
        \includegraphics[width=0.9\textwidth]{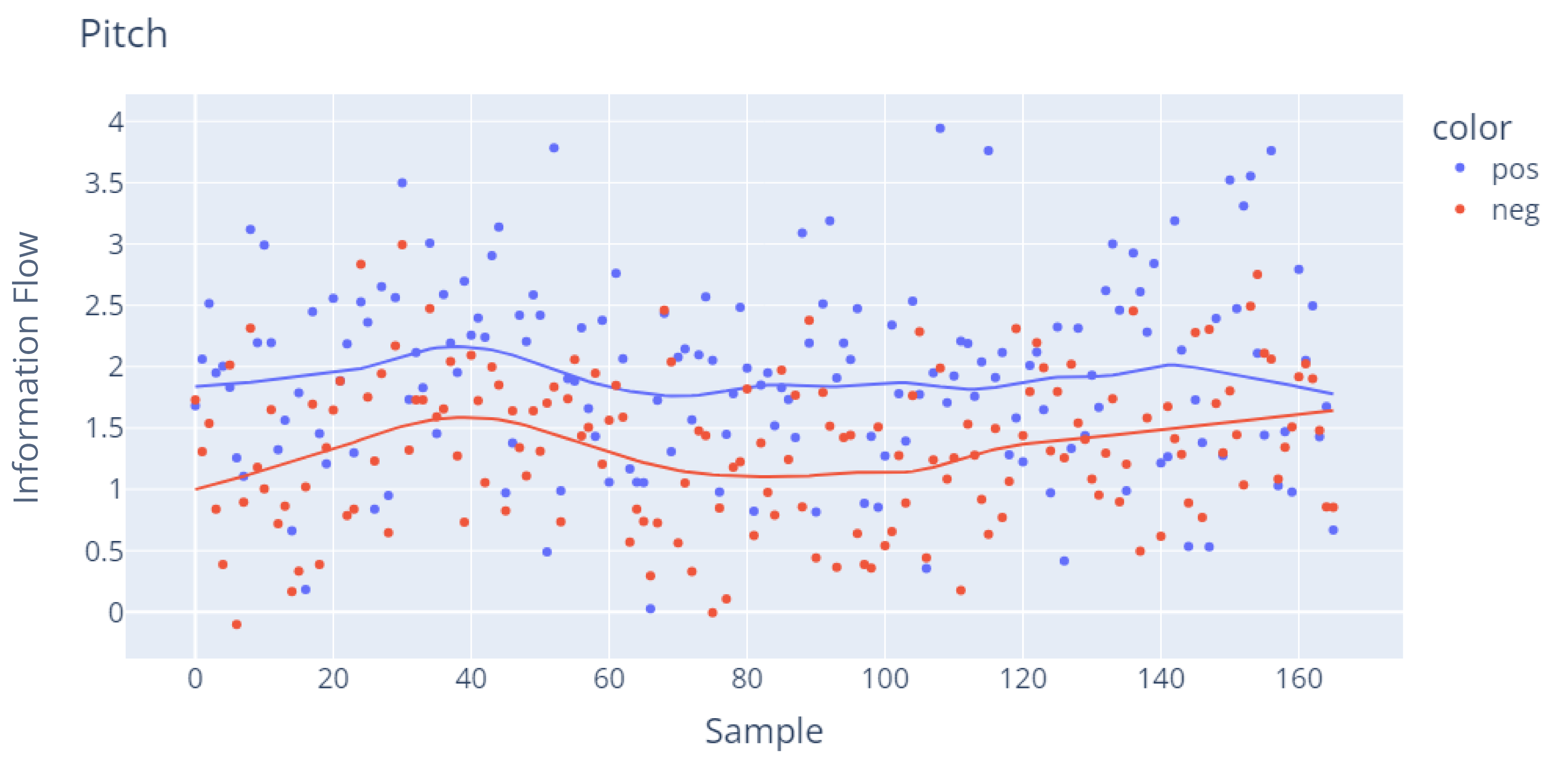}
        \caption{}
        \label{fig:second}
    \end{subfigure}
    \begin{subfigure}[b]{0.5\textwidth}
        \centering
        \includegraphics[width=0.9\textwidth]{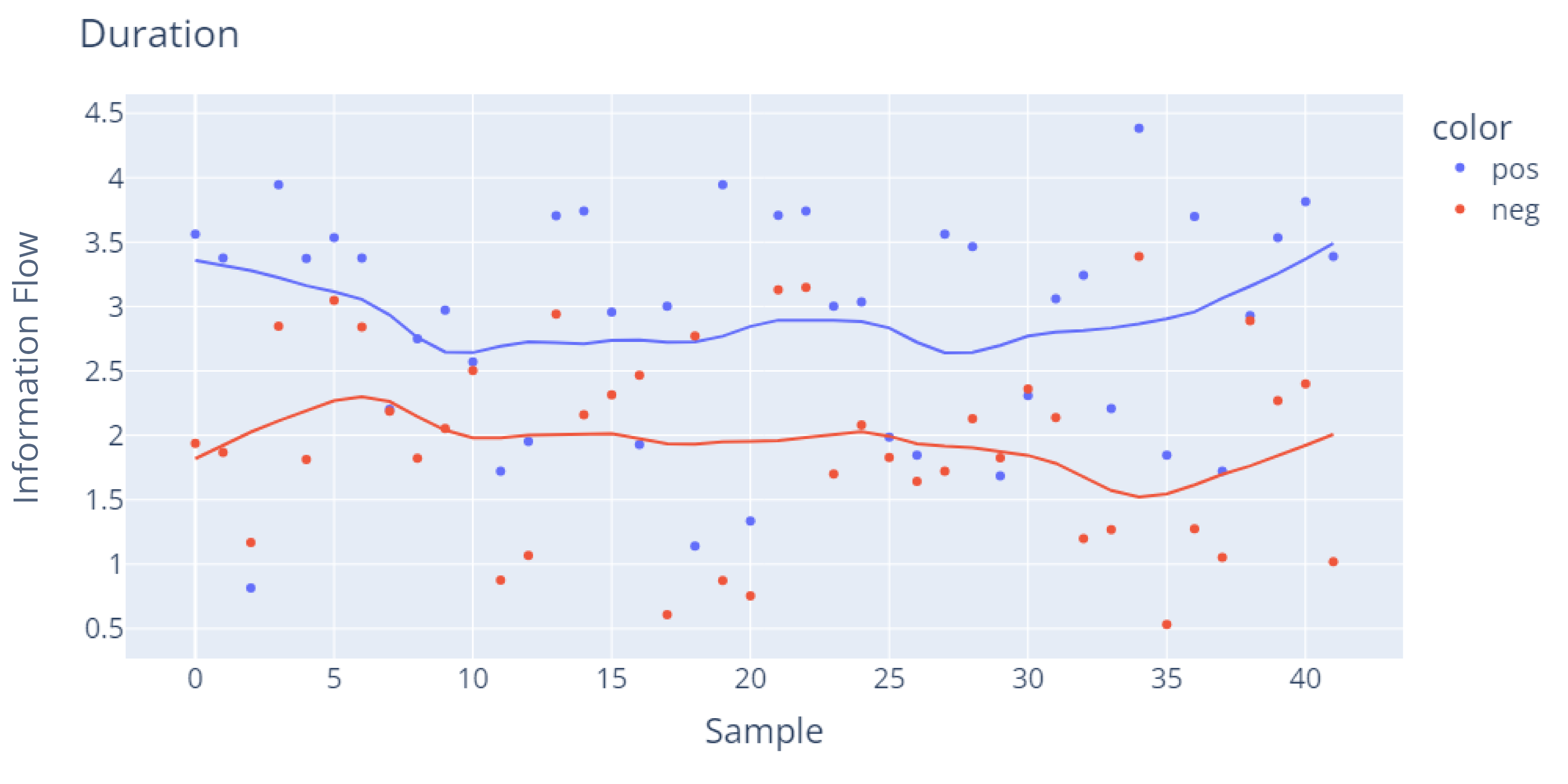}
        \caption{}
        \label{fig:third}
    \end{subfigure}%
    \begin{subfigure}[b]{0.5\textwidth}
        \centering
        \includegraphics[width=0.9\textwidth]{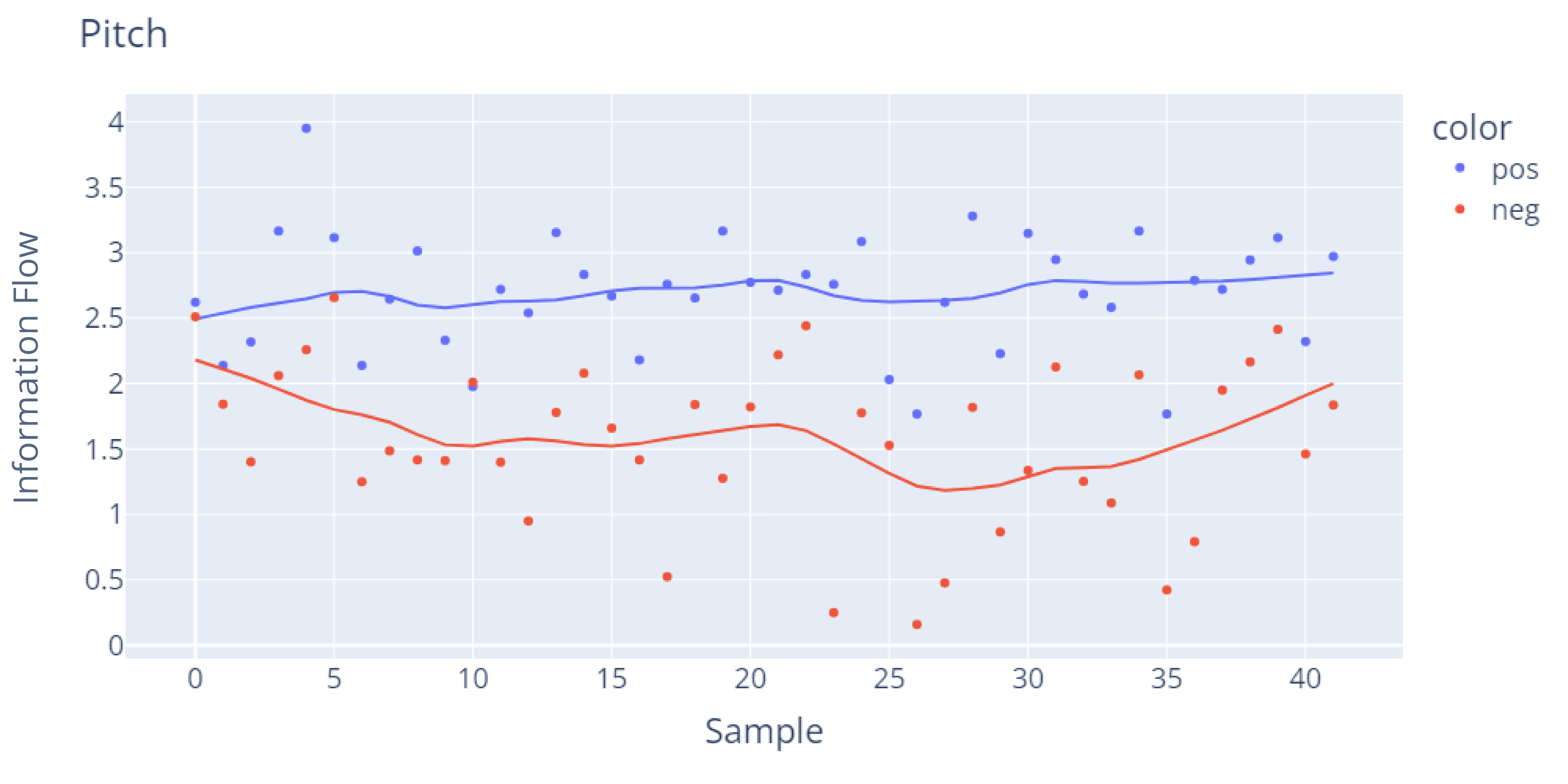}
        \caption{}
        \label{fig:fourth}
    \end{subfigure}
    \caption{(a,b) Comparison of information flow with respect to duration and pitch in the music for the Score dataset. (c,d) Comparison of information flow with respect to duration and pitch in the music for the URMP dataset. We see that for both the datasets and both pitch and duration, the information flow in the positive samples is more than negative samples.}
    \label{fig:info_flow}
\end{figure*}

\subsubsection{Data Representation}
\label{datarep}

We use data in the MIDI format, which is converted to the data representation from \cite{mmt} required by the transformer. The representation is able to represent two to three times longer music samples for the same sequence length than popular representations like \cite{figaro,mmm} used for multitrack music generation.  Each music sample is represented as a sequence of events, and each event $x$ is represented as a tuple of six variables:
\begin{equation*}
    x = (x^{type}, x^{beat}, x^{position}, x^{pitch}, x^{duration}, x^{instrument})
\end{equation*}
$x^{type}$ indicates the type of the event. Table \ref{tab:type} shows the five values it can take and their meanings. 
\begin{table}[H]
    \centering
    \renewcommand{\arraystretch}{1.4}
    \begin{tabular}{| >{\raggedleft}m{0.8cm} | m{6cm} |}
    \hline
    $\mathbf{x^{type}}$  & \textbf{Event explanation} \\
    \hline
    0  & indicates the start of the music sample \\
    \hline
    1  & indicates an instrument used in the music sample \\
    \hline
    2  & indicates the end of the instrument list and the start of notes  \\
    \hline
    3  & indicates a note  \\
    \hline
    4  & indicates the end of the music sample \\
    \hline
    \end{tabular}
    \caption{The five different types of events and what they represent}
    \label{tab:type}
\end{table}
For start-of-music ($x^{type} = 0$), start-of-notes ($x^{type} = 2$), and end-of-music ($x^{type} = 4$) events, the remaining five variables are set to zero. Instrument events ($x^{type} = 1$) appear after the start-of-music event ($x^{type} = 0$) and before the start-of-notes event ($x^{type} = 2$) and act as a list of the instruments used in the musical piece. For an instrument event, the variable $x^{instrument}$ indicates the MIDI program number of the instrument, and the remaining four variables are set to zero. 
For a note event ($x^{type} = 3$), the remaining variables define the onset, pitch, duration, and instrument of the note. $x^{beat}$ denotes the index of the beat that the note lies in, and $x^{position}$ denotes the position of the note within the beat. The onset of the note would be $ x^{beat} \cdot r + x^{position}$, where $r$ is the temporal resolution of a beat. 

\subsubsection{Workflow}
Algorithm \ref{alg_1} explains the overall workflow of our method. Given a MIDI file with two tracks, we first split the tracks to obtain the individual tracks $X$ and $Y$. Then for each of the individual tracks $X$, $Y$ and the combined track $XY$, we compute the conditional entropy given its past, i.e $H(X|\bar{X})$, $H(Y|\bar{Y})$ and $H(XY|\overline{XY})$ using equation \ref{cond_entropy}. To compute the expectation over time, we start with a small segment of the music as our starting context. The $start$ variable in the algorithm is a sliding window that feeds the MTMT.Transformer (Multitrack Music Transformer) with the conditional sequence (past of the variable). For example, to calculate $H(X|\bar{X})$, we slide over the past of sequence $X$ with some context window. Then we obtain the probability distribution over the next output token. Using this probability, we compute the entropy at that timestep. We obtain the conditional entropy $H(X|\bar{X})$ by averaging the computed entropies over time. Similarly, we compute $H(Y|\bar{Y})$ and $H(XY|\overline{XY})$ to calculate the total information flow. Intuitively, the transformer model (MTMT.Transformer) should provide low entropy distributions if knowing the past sequence makes it easier to predict the next token.

\section{Experiments}
We conduct in-depth objective and subjective experiments to evaluate the effectiveness of our score. Below, we first introduce the datasets we use, our results, and the studies we conducted.

\subsection{Datasets}

We conduct our experiments using two datasets:
\begin{enumerate}
    \item A dataset we curated using music scores from the website MuseScore. We will refer to this dataset as the Score dataset.
    \item The University of Rochester Multi-modal Music Performance (URMP) Dataset \cite{umrp}.
\end{enumerate}
\begin{figure*}
    \centering
    \includegraphics[width=\textwidth]{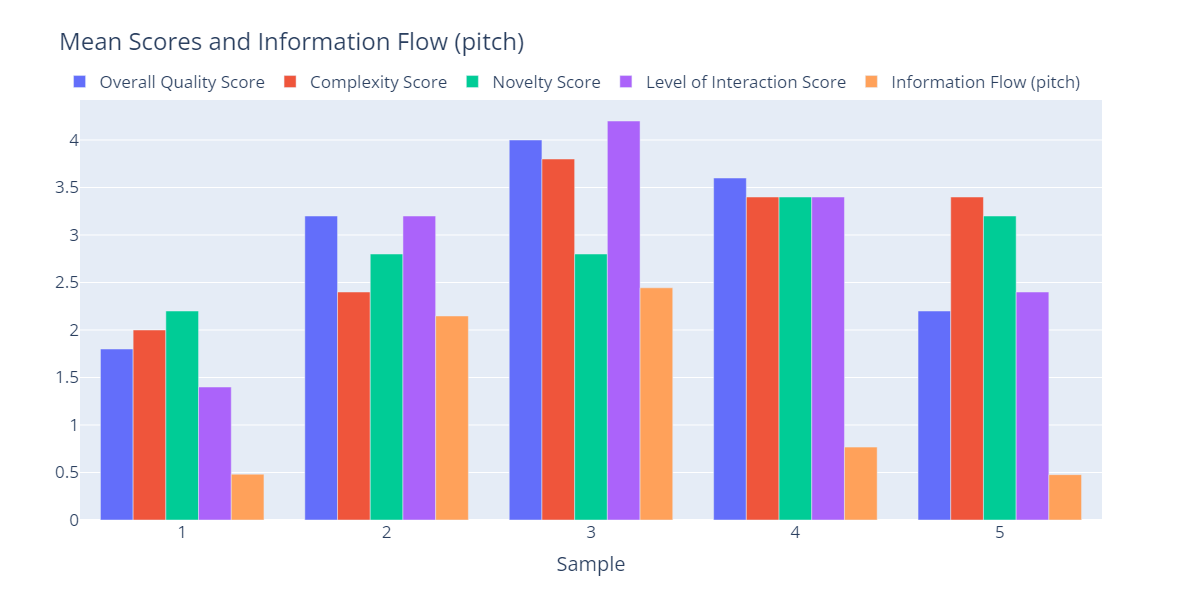}
    \caption{We compare the information flow (for the pitch variable) with the scores obtained from the qualitative study.}
    \label{fig:qualitative_study}
\end{figure*}
\subsubsection{Score dataset}
The dataset consists of 500 musical pieces from MuseScore. Each piece has two tracks - a piano melody and its piano accompaniment. The MuseScore files are converted to the MIDI format using the MusPy Python library \cite{muspy}. We pick 200 samples from this dataset as positive (matching) melody-accompaniment pairs for our experiments. We then replace the accompaniment tracks in them by picking tracks uniformly at random from the remaining samples to form corresponding negative melody-accompaniment pairs. The results shown have less than 200 samples because we removed some samples that had problems while converting to the data representation format described in \ref{datarep}. 

\subsubsection{URMP dataset}
The URMP dataset \cite{umrp} comprises 44 simple multi-instrument musical pieces constructed from musically coordinated and separately recorded performances of individual tracks. The dataset consists of tracks made using thirteen different instruments. We use only the musical scores provided in the MIDI format. The dataset also contains high-quality individual instrument audio recordings and videos of the pieces. We removed two samples from this dataset for our experiments due to problems while converting to the data representation format described in \ref{datarep}. We created 42 positive melody-accompaniment pairs and 42 corresponding negative melody-accompaniment pairs using the same uniform random selection method followed for the Score dataset.

\subsection{Analysis}
We demonstrate the effectiveness of our method in identifying ``good" music in Figure \ref{fig:info_flow}. Ideally, musically pleasing pieces should have a higher score, while non-coherent music (our negative samples) should have a lower score. As shown in Figure \ref{fig:info_flow}, the overall score trend for positive samples is considerably higher than for negative samples. Since MTMT uses a six-tuple representation, we can compute our score for each variable to further understand the creativity process involved. For example, we can analyze how much information about pitch flows from one track to another by computing the information flow only for the pitch variable.

\subsubsection{Qualitative Study} 
In order to evaluate if our method also aligns with human perception of ``good" and creativity, we conducted a qualitative study. We chose five music samples that included both positive and negative samples. We asked 5 expert musicians in our social circle to evaluate each musical sample on the overall quality of the music, novelty, complexity, and level of interaction in the music. Each criterion was given a score out of five. We then compare the information flow with the survey responses. Figure \ref{fig:qualitative_study} shows the overall trend for our score and the survey responses for the five samples. Overall, information flow aligns with human perception of quality. Samples 2 and 3 are positive samples, and we see a higher quality score and information flow. We also observe a higher score for the Level of Interaction metric. This denotes how much musical interaction is present between the two tracks in the music sample. We notice a high level of interaction score for positive samples, while it is relatively low for negative samples (1 and 5).

\begin{figure*}[t] 
    \centering
    \begin{subfigure}[b]{0.5\textwidth} 
        \centering
        \includegraphics[width=\textwidth]{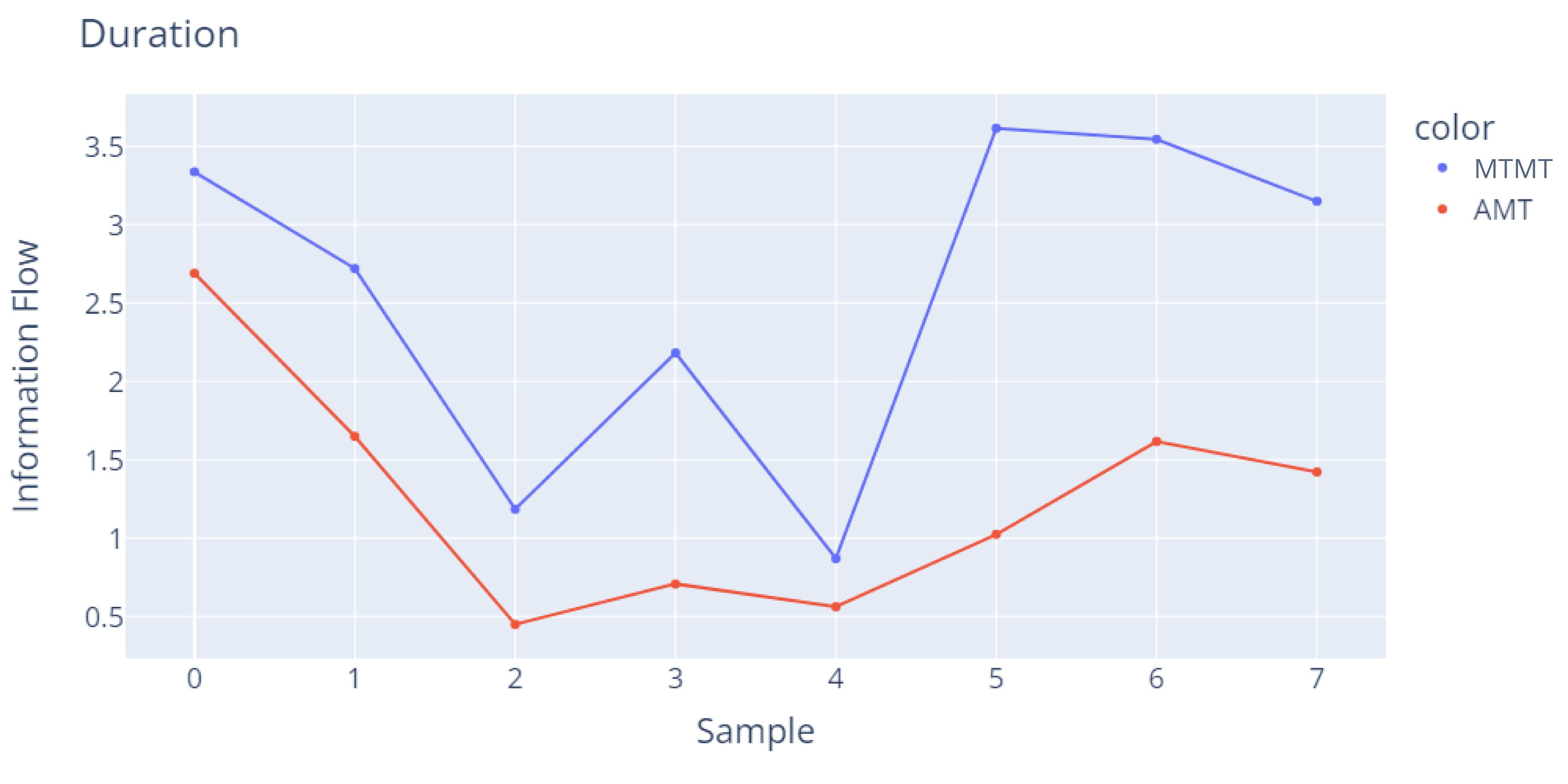} 
        \caption{}
        \label{fig:sub1}
    \end{subfigure}%
    ~ 
    \begin{subfigure}[b]{0.5\textwidth} 
        \centering
        \includegraphics[width=\textwidth]{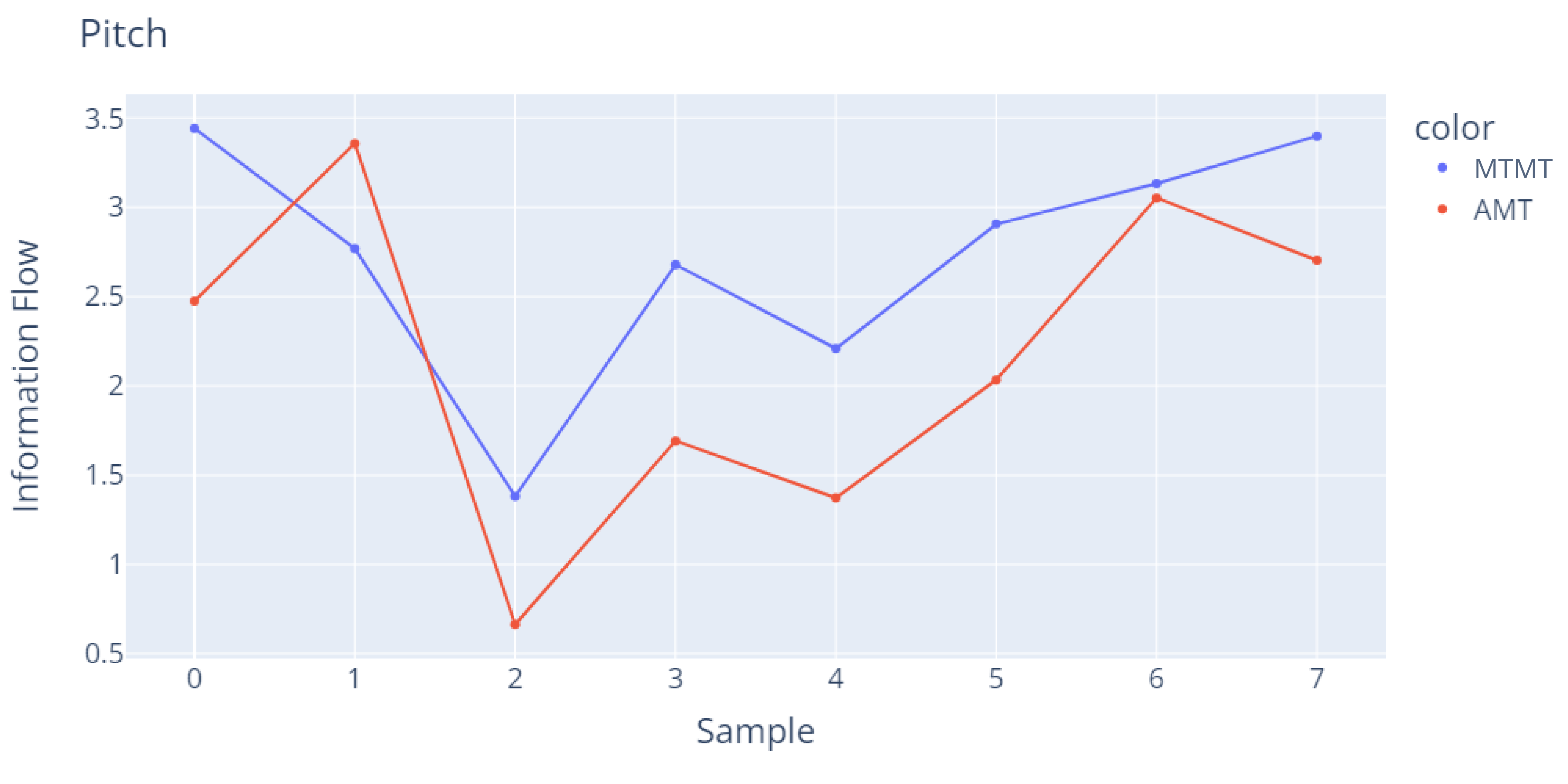} 
        \caption{}
        \label{fig:sub2}
    \end{subfigure}
    \caption{Self-Enhancement Bias: (a) Comparison of information flow with respect to duration on MTMT vs AMT generated sequences, (b) Comparison of information flow with respect to pitch. We observe that the self-enhancement bias observed in popular LLMs is also observed when using MTMT as the entropy estimator.}
    \label{fig:test}
\end{figure*}

Interestingly, we found that many participants rated sample 4 (a negative sample) with high overall quality and level of interaction scores. Although we created sample 4 by randomly mixing two tracks from separate music pieces, a lot of participants seemed to have preferred the combined track. This particular example (example 4) happens to fit musically together since one of the tracks is repetitive in a high register, with the other playing a repeated melodic pattern in a lower register (ostinato), thus establishing a nice separation with minimal clashes in terms of rhythm or underlying harmony. This happy coincidence results in a mashup that is perceived as musically pleasing and the voices meaningfully interacting.
This also demonstrates the difficulty in understanding the creative process in humans. Looking at the novelty score, we can infer that participants found some of these negative samples as novel and something new, but rated them as low quality (except for sample 4). Note that none of our samples in the survey were ``simple" toy examples. They were all complex musical pieces, as indicated by the high complexity score across all the samples. 

\subsubsection{Positional Bias}
Several popular transformer-based large language models (LLMs) have been shown to suffer from positional bias \cite{zheng2023judging}. Positional bias refers to the fact that a model comes to a different conclusion based on the position of the input. For example, \cite{zheng2023judging} show that in a task of evaluating which sentence is well structured, GPT-4 has been shown to favor one sentence over another depending on which comes first. Since we are using transformer-based architectures to compute our score, it is essential to evaluate our method for such biases.

We conduct an experiment to verify if our method has positional bias. Ideally, the score should remain the same for both orderings of tracks $XY$ and $YX$ as they are the same music. However, since the transformer can output different probabilities based on which track is first, there might be positional biases. To confirm if this is true, we compare our scores with and without reversing the tracks. We consider 166 samples for this experiment and report the Mean Squared Error (MSE) between both the orderings of the tracks (XY and YX). Table \ref{tab:positional} shows that with MTMT, the entropy estimator does not suffer from positional bias as the MSE is close to zero.
\begin{table}[h]
    \renewcommand{\arraystretch}{1.4}
    \centering
    \begin{tabular}{|c|c|c|c|}
        \hline
        & Duration & Pitch & Position \\
        \hline
        MSE & 0.02 & 0.003 & 0.001 \\
        \hline
    \end{tabular}
    \caption{Evaluating Positional Bias: We find that the position/order of the tracks does not affect the entropy estimators.}
    \label{tab:positional}
\end{table}

\subsubsection{Self-Enhancement Bias}
Another bias in transformer-based architecture is the self-enhancement bias, i.e. models tend to prefer their own generations versus other model creations \cite{zheng2023judging}. For example, it is known that GPT-4 prefers text generated by GPT-4 compared to text generated by other LLMs. This is an issue, as we want our score to reflect the quality of the music irrespective of what generating process was involved in creating the music. For example, since we are using MTMT, our method should not give a higher score to musical compositions from MTMT compared to another musical piece.

To test if self-enhancement bias exists in our method, we compute our score on similar generated music from both MTMT and the Anticipatory Music Transformer (AMT) \cite{anticipatory}. AMT is a symbolic music transformer based on the proposed method of anticipation. The model can perform infilling and generation of multitrack symbolic music. For both models, we feed in the same starting context and ask the model to continue generation for an equal number of timesteps. Now, we have similar music from two different generative models. Note that both the models have different training procedures, so the generated music might differ vastly in spite of having the same starting sequence. Figure \ref{fig:test} shows the information flow scores for eight samples from our experiments. Interestingly, we observe that the bias observed in most LLMs is also observed in MTMT and our framework. The model tends to prefer its own compositions more. Theoretically, from equation \ref{cond_entropy}, we see that for information flow to be higher, $H(XY|\overline{XY})$ needs to be low. When computing this term on its own generations, the model is easily able to predict the next token and hence outputs a low entropy value, causing an increase in the overall information flow. While we use data that was not part of MTMT's training process in our other experiments, it is important to understand the effects of such biases in our method.

\section{Future Work}
We present a method to compute information flow and use it as a measure to evaluate creativity in AI-generated music. In the future, we would like to come up with optimization objectives based on this score. Ideally, we should be able to guide a model to output music such that the information flow would be maximized. Moreover, we would like to apply our framework in evaluating creative processes in other domains such as text in conversational systems, video-to-video generation systems for applications such as dance, etc. Our method is data and application-agnostic and only needs a pre-trained generative model to estimate the entropy. This method could also have huge applications in the analysis of financial time series.



\section*{Acknowledgments}
This project has received funding from the European Research Council (ERC) under the European Union’s Horizon 2020 research and innovation programme (Grant agreement \#883313, project REACH : Raising Cocreativity in CyberHuman Musicianship)


\bibliographystyle{named}
\bibliography{ijcai24}

\end{document}